\documentclass[twocolumn,secnumarabic,amssymb, nobibnotes, aps, prl]{revtex4-1}

\usepackage{graphicx}
\usepackage{subfigure}
\usepackage{epsfig}
\usepackage{epstopdf}
\usepackage{natbib}
\usepackage{amsmath}
\usepackage{braket}
\usepackage{amsmath,amssymb}
\begin{document}
\title{Generation of non-classical states of photons from metal-dielectric interface: a novel architecture for quantum information processing}%
\author{Karun Mehta}  
\email{karun.mehta@iitrpr.ac.in}
\author{Shubhrangshu Dasgupta}
\affiliation{Department of Physics, Indian Institute of Technology Ropar, Rupnagar, Punjab 140001, India}
\date{\today}
\begin{abstract}
We show that it is possible to generate photons in nonclassical states from a metal-dielectric interface using quantum emitters on the interface. The photons emitted into the surface plasmon mode from the initially excited emitters radiate out in free space in a cone-shaped geometry. When detected at two detectors, these photons exhibit anti-coalescence, a clear signature of nonclassicality. Such a system can also be employed as a building block for a distributed quantum network. We further show that it is indeed feasible to implement our model using available technology.
\end{abstract}
\maketitle

In recent times, there has been substantial development in techniques of long distance quantum communication and distributed quantum computing. Photons with nonclassical properties play the key role in this aspect, as such properties are carried to a distant node of a distributed quantum network via photons, either through a fiber or the free space. In addition, while the major advances in the area of quantum computing have been made in different kinds of architecture, namely, atoms interacting with cavity fields (``cavity QED") \cite{haroche,saffman,rempe}, trapped ions \cite{Blatt}, nuclear magnetic resonance \cite{Vandersypen}, photons \cite{kok}, quantum dots \cite{loss}, and superconducting systems \cite{Ze}, they are limited to the systems with a few particles or qubits. This also limits the number of photons that would be required for a scalable communication. In this context, the condensed matter system (an array of spins, for example) seems to be quite promising to provide suitable scalability, while the photons are preferred to for building up communication at a large distance.

 In this Letter, we develop a suitable platform, based on a metal-dielectric interface, that combines the best of both the architectures - in terms of scalability as well as long-distance communication. We will particularly focus on how to generate photons in a certain class of nonclassical states, based on interaction of quantum emitters with surface-bound electromagnetic field modes on such interface. These photons can be used for long-distance communication through free space or fiber. More importantly, the method, we describe next, is scalable to a large number of photons.

On a metal-dielectric interface, there exists a transverse magnetic (TM) mode of electromagnetic field (as allowed through the boundary conditions across the interface), that propagates as a plane wave along the $x$-axis (see Fig. 1). When coupled with the plasma oscillations of the electrons on the metal surface, one obtains certain quasi-modes, referred to as surface plasmons (SP) \cite{Maier}. The wave in the SP mode is evanescently confined in the perpendicular direction (i.e., into both the media in $\pm$ve $z$-directions) \cite{Raether}.

To excite a localized SP mode, one can illuminate the metal surface with single photons \cite{Tame}. An alternative way is to use a quantum emitter, emission from which can also excite an SP mode. For example, localized plasmons have been excited using single molecules \cite{anger}. The excitation of a propagating SP mode (instead of a localised one) using a quantum dot (QD) has been first reported in \cite{akimov}. The fluorescence from a CdSe quantum dot is coupled to a nanowire, and the light scattered from the end of the nanowire displayed an anticoalescence behaviour. In presence of many atoms, the correlation in multiatom fluorescence, that couples to the guided mode of a nanofiber, has been studied in detail \cite{hakuta}, for several atomic arrangements.

Various nonclassical effects in plasmonic setups have also been observed. Possibility of photon coalescence using plasmonic waveguides \cite{coa,fujii} (akin to the Hong-Ou-Mandel experiment using light \cite{hong}) and anticoalescence using plasmonic beamsplitters \cite{vest} have been demonstrated. Path entanglement between two photons are shown to retain when these photons were converted into surface plasmons by putting an SP waveguide in both the arms of a Mach-Zehnder interferometric setup \cite{fakonas}. Even a single photon can be entangled with a single SP mode \cite{dheur}, that enables one to remotely prepare a single-plasmon state. It is further shown that squeezing in photons can be mapped into SP modes \cite{huck}, paving the way to plasmonic sensors.

In this Letter, on the other hand, we show how to generate, from the metal-dielectric interface, the photons in their nonclassical states. The two quantum emitters (say, quantum dots) are placed at a certain distance from the interface (inside the dielectric), which radiatively couple to the SP mode. The photons emitted by these emitters propagate along the interface and radiate out into the free space \cite{maradudin}. The nonclassicality in these photons is inherent, thanks to the correlation between the emitters. We emphasize that our setup is different from the previous works \cite{akimov,hakuta} in which the emitters get coupled to either propagating SP mode in a nanowire or the guided mode of the nanofiber, instead of {\it the propagating SP mode on the interface}.

Let us start with two identical quantum dots (QDs), placed very close to a metal-dielectric interface at a fixed distance $z_0 (>0)$. Each QD can be considered as a two-level system with the relevant energy levels $\ket{g}$ and $\ket{e}$, the transition frequency $\omega_{0}$, and the dipole moment $\vec{\mu}$. These QDs can couple to the SP modes on metal-dielectric interface.
Two detectors D1 and D2 are placed in the far-field region at the end of the interface to detect the photons.
\begin{figure}[h]
\centering
\includegraphics[scale=0.24]{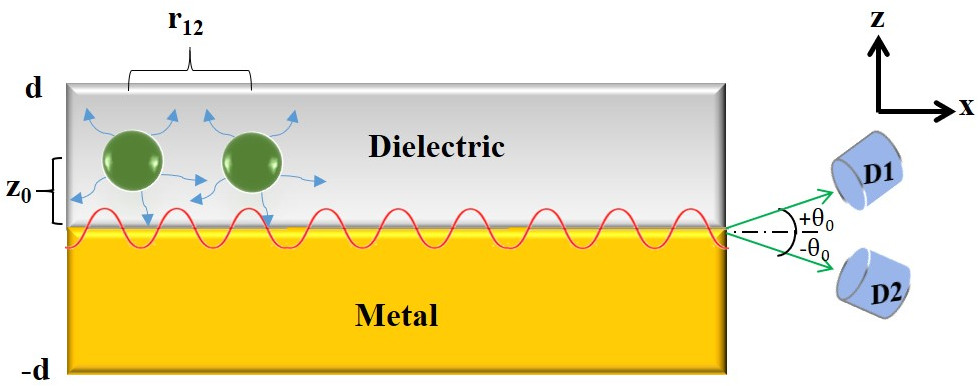}
\caption{Radiative coupling of QDs to the SP.}
\end{figure}

Note that the coupling of the QDs with the SP mode depends on the location and orientation of the QDs \cite{moreno}. We assume that both the dots have the dipole moments oriented in the same direction, and are initially in the respective excited state.
The decay rate of the QDs gets modified to a value larger than that in free space, as the density of states in the vicinity of the metal surface increases. The total decay rate $\gamma$ of each QD can be contributed from different decay channels, as
\begin{equation}
\gamma=\gamma_{\rm rad} + \gamma_{\rm non-rad} + \gamma_{\rm SP},
\label{gamma}
\end{equation}
where $\gamma_{SP}$ is the rate of decay into the SP mode, and $\gamma_{rad}$ ($\gamma_{non-rad}$) is that into radiative modes (various non-radiative modes including dissipation in metal). For large distance from the interface, radiation into free space dominates over other channels, while the decay rate into the SP mode decreases exponentially with $z$. At a moderate distance from the interface, the non-radiative process has only a negligible effect due to $z^{-3}$ dependence and the decay of the QDs is dominated primarily into the SP mode \cite{sambles}.

We consider that the dots are placed at a distance $r_{12}$ from each other along the interface, such that $r_{12}$ is much less than the wavelength $\lambda_{SP}$ of the SP mode. If the dots emit photons resonantly with the SP mode, then these two quantum dots can coherently couple to the single surface plasmon mode. Our particular interest is in the excitation of the SP mode while the other channels are treated as dissipation mechanisms. The photons in the SP mode will propagate along the metal-dielectric interface in +ve $x$-direction, so that at the end of interface they scatter out into free space mode \cite{akimov,maradudin} and get detected by the detectors D1 and D2. We assume here that the interface is smooth enough so that the photons do not decay along the direction of propagation and the distance they travel till the end of the interface is much smaller than the propagation length.
The Hamiltonian of the joint system comprising the dots and the SP mode can be written as \cite{zubairy}
\begin{equation}
\mathcal{H}=\mathcal{H}_{0}+\mathcal{H}_{dd}+\mathcal{H}_{SP} \;,
\label{totalH}
\end{equation}
where
\begin{eqnarray}
\mathcal{H}_{0}&=&\frac{\hbar}{2}\omega_{0}\sum_{i=1}^{2} (S_i^+ S_i^- - S_i^- S_i^+)\;,\\
\mathcal{H}_{dd}&=&\hbar \Omega_{12} (S_1^+ S_2^- + S_2^+ S_1^-)\;,\\
\mathcal{H}_{SP}&=&\frac{\hbar}{2}\sum_{i=1}^{2}(\Omega_{i}S_i^+e^{-\iota\omega t}+{\rm h.c})\;.
\end{eqnarray}
Here $S_i^+$ and  $S_i^-$ are the energy raising and lowering operators for the $i$th dot with the transition frequency $\omega_0$, $\hbar\Omega_{12}$ is dipole-dipole interaction energy, representing direct interaction of the two dots, and $\Omega_i$ is the Rabi frequency of the $i$th dot, when driven by the electromagnetic field $\mathbf{E}$ in the SP mode. This can be written as
$\Omega_{i}=\mathbf{d_i.E}/\hbar$, where the field $\mathbf{E}$ is given by
$\mathbf{E}=E_x\left[1,0,\frac{-k_x}{k_{z_1}}\right]e^{\iota(k_xx+k_{z_1}z)}$. Here the wave vector of the field along $x$-direction is given by
$k_x=\sqrt{\frac{\varepsilon_1 \varepsilon_2}{\varepsilon_1+\varepsilon_2}}$, where $\varepsilon_1$ and $\varepsilon_2$ are the permittivities of the metal $(z<0)$ and the dielectric $(z>0)$, respectively. The wave-vector $k_{z_j}=\sqrt{\varepsilon_j k_0^2-k_x^2}$ ($k_0$ being the wave vector in free space) in the $z$-direction is imaginary in both the media and represents a decaying amplitude along $z$-direction (i.e., perpendicular to the interface). Here we emphasize that the SP mode is treated classically, unlike in \cite{moreno,alex,susa}, which is justified as the mode is freely propagating (not confined) plane wave along $x$-direction, at a fixed value of $z$.

\begin{figure}[h]
\centering
\includegraphics[scale=0.37]{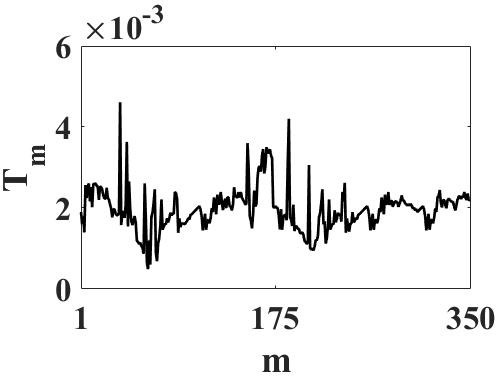}
\caption{Transmissivities of the surface plasmons into radiative modes. $\epsilon_1=-5.65+0.65\iota $ and  $\epsilon_2=1, \lambda_0=450$ nm, $\lambda_{SP}=0.91\lambda_0$. The total transmission into radiative modes is 68\% of the SP mode across all possible modes.}
\label{tm}
\end{figure}

The photons in the SP mode propagate along the metal-dielectric interface and at the end of interface $(x=0, z=0)$, they scatter into free space (i.e., radiative) modes, however, with a finite probability of transmission across the interface boundary. The corresponding transmission coefficients $T_m$ into the $m$th radiative mode have been evaluated using boundary conditions for discrete equally spaced points along the interface $x=0$ \cite{maradudin} (see Supplementary Material for calculation of $T_m$). In Fig. \ref{tm}, we show the distribution of the transitivity across a number of radiative modes. We further find that the transmission probability into the free space modes is quite large, if the permittivity $\epsilon_2$ of the dielectric is close to unity. Henceforth, in the rest of this paper, we choose an air-metal interface, such that $\epsilon_2=1$. In such a configuration, it would also become convenient to use lasers to initially excite the dots.

At the end of the interface, the angular distribution of the scattered intensity in the far-field is given by \cite{hu}.
\begin{equation}
I(\theta)\varpropto\frac{\cos^2\theta}{\lambda_{SP}}\left|\int_{-\infty}^{+\infty}dz H_{y,{SP}}e^{\iota k_0z \sin\theta}\right|^2
\end{equation}
where $H_{y,{SP}}$ is the $y$-component of the magnetic field in the SP mode and $\theta$ is the angle with the +ve $x$-axis. Interestingly, we obtain two values of $\theta$ (say, $\pm \theta_0$ in Fig. 3) at which the intensity is maximum. The lower values of $m$ for which the $T_m$ is dominant, contribute the most to the intensity. This suggests that in the $x-z$ plane, there exists two possible directions at which the photons can be detected, with maximum probability. We consider two detectors D1 and D2 at such configuration.

\begin{figure}[h]
\centering
\includegraphics[scale=0.34]{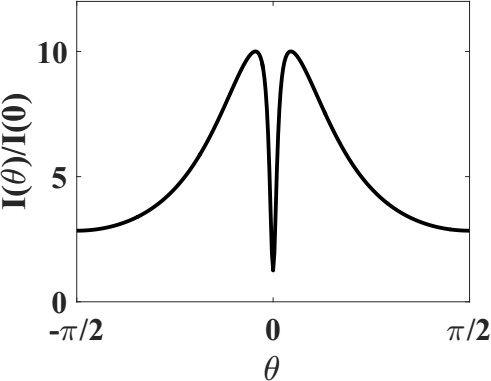}
\caption{Far-field angular distribution of the SP field scattered at the end of the interface for $\epsilon_1=-5.65+0.65\iota $ and  $\epsilon_2=1, \lambda_0=450$ nm, $\lambda_{SP}=0.91\lambda_0$. Here $\theta_0$ is found to be 0.13 radians. }
\end{figure}
As the two dots interact directly with each other and cross-talk through their common interaction with the same SP mode, there would exist a correlation between the photons emitted by them. Such a correlation between these photons could be indicated when detected at the detectors D1 at position $\mathbf{R_1}$ and  D2 at $\mathbf{R_2}$ in the far-field regime. These two detectors would only detect the number of photons, irrespective of their polarization. We have calculated the intensity-intensity correlation of the field emitted from both the dots, at the end of interface by placing two detectors as shown in Fig. 1. The correlation function of the electromagnetic field can be represented in terms of the second-order correlation function \cite{zubairy} of spin operators of each two-level dot, as
\begin{multline}
g^{(2)}(\mathbf{R_1,R_2};\tau)=\lim_{t\to\infty}\sum_{p,q,r,s=1}^{2}e^{\iota k(\hat{R}_{1}.\vec{r}_{ps}+\hat{R}_{2}.\vec{r}_{qr})}\\\times
\frac{\langle S^+_p(t)S^+_q(t+\tau)S^-_r(t+\tau)S^-_s(t)\rangle}{\left[\sum_{m,n=1}^{2} \langle S^+_m(t)S^-_n(t) \rangle e^{ik\hat{R}.\vec{r}_{mn}}\right]^2}
\label{g2}
\end{multline}
where $\tau$ is the time-delay between two photons when detected at the detectors D1 and D2.

Here, we have calculated the expectation values of the higher-order combinations of spin operators using the quantum regression theorem \cite{carmichael} and the solutions of the following master equation of the density matrix $\rho$ of the two dots:
\begin{equation}
\frac{\partial \rho}{\partial t}=\frac{1}{\iota\hbar}[H,\rho]-\sum_{i,j=1}^{2}\gamma([S^+_i,S^-_j\rho]-[S^-_j,\rho S^+_i]\;,
\label{master}
\end{equation}
where $H$ is given by the Eq. (\ref{totalH}) and $\gamma$ is enhanced decay rate, as given in Eq. (\ref{gamma}).

\begin{figure}[h]
\centering
\includegraphics[scale=0.35]{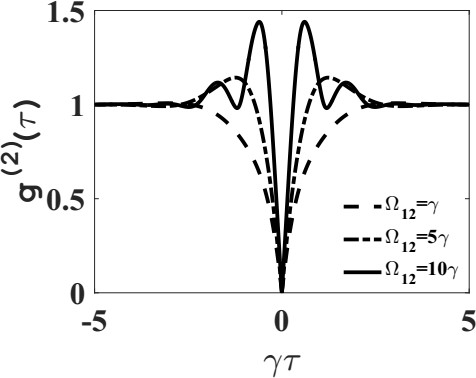}
\caption{Second order correlation function for different interaction energy between the QDs. $\Omega_1=\Omega_2=\gamma, \gamma=2.9\times10^{10}$ Hz.}
\label{g2fig}
\end{figure}

In Fig. \ref{g2fig}, we display the behavior of the second-order correlation function $g^{(2)}(\tau)$. We found that $g^{(2)}(0)=0$, which signifies that both the detectors D1 and D2 do not click simultaneously - in other words, both the photons reach at either detector at the same time, but not at two different detectors. This is a clear signature of anticoalescence of photons and therefore a sub-Poissonian statistics, indicating nonclassical behaviour of the photons. This is a situation akin to Hanbury Brown-Twiss experiment \cite{hbt}, where the role of the beam-splitter is played by the end of the interface. Moreover, such a behaviour persists for all values of dipole-dipole coupling strength $\Omega_{12}$, which is a function of $r_{12}$, the distance between the dots. This signifies that the time of emission of the photons from the dots and the path difference $r_{12}$ while reaching at the detectors do not play any role in the photon statistics.

For weaker interaction (small $\Omega_{12}$), when the QDs are far apart, we find that $g^{(2)}(\tau\ne 0)$ increases monotonically from zero to 1. On the other hand, for larger $\Omega_{12}$ (for closer positions of the QDs), the correlation function oscillates with $\tau$, that arises due to Rabi coupling between the dots via sharing of emitted photons. For very large $\tau$, however the non-classical correlation ceases to exist as $g^{(2)}(\tau)$ approaches to 1.
\begin{figure}[h]
\centering
\includegraphics[scale=0.33]{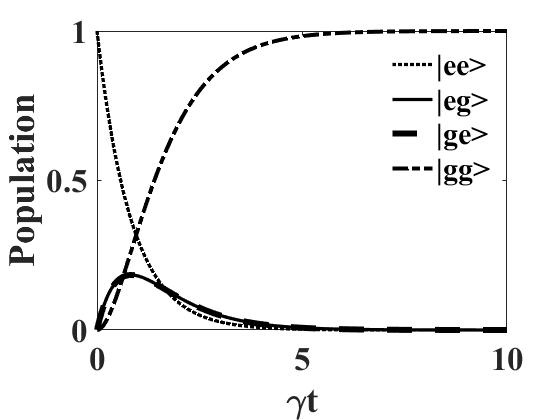}
\caption{Dynamics of population of the joint states $|ee\rangle$, $|eg\rangle$, $|ge\rangle$, and $|gg\rangle$ of the two dots.  $\Omega_{12}=\gamma, \Omega_1=\Omega_2=\gamma,  \gamma=2.9\times10^{10}$ Hz.}
\label{pop}
\end{figure}

Our proposed model can be realized on a silver-air interface where dielectric constant for silver is $\epsilon_1=-5.65+0.65\iota $ at wavelength $\lambda_0=450$\ nm $ (\lambda_{SP}=0.91\lambda_0)$ \cite{johnson} and that for air $\epsilon_2=1$, with the corresponding propagation length of surface plasmon mode $\sim16\ \mu $m  and the evanescent decay of the SP mode along $z$-direction is $\sim180\ $nm into the dielectric \cite{Maier}.  One can maintain a constant distance $z_0$ between the dots and the interface using a few nanometer spacer layer \cite{moreno} with the dielectric constant close to unity. We choose $\gamma=1.2\gamma_{0}, \gamma_{0}$ being the spontaneous decay rate into radiative modes of the free space.

The anticoalescence behaviour indicates that the photons are detected at the states $|2,0\rangle$ or $|0,2\rangle$ with equal probability, where $|n,m\rangle$ refers to a situation with $n$ photons detected at the detector D1 and $m$ photons at D2. This is irrespective of the radiative modes that the photons have been emitted into. This, therefore, indicates formation of a path-entangled state of the two photons, that can be written as $|\psi\rangle_2\equiv (|2,0\rangle + e^{\iota \phi}|0,2\rangle)/\sqrt{2}$. The relative phase $\phi$ is washed out upon detection; however, it could be measured using homodyne detection techniques. That the two photons are indeed emitted by the dots has also been verified by solving the master equation (\ref{master}) with both the dots initially excited. We have found that the probability that both the dots decay to the ground states is unity at the steady state (see Fig. \ref{pop}). This indicates that these dots must have emitted two photons.

\begin{figure}[h]
\centering
\includegraphics[scale=0.35]{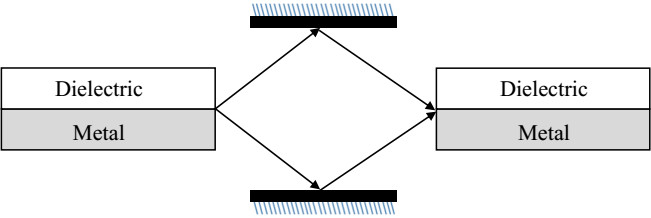}
\caption{A possible architecture for long-distance quantum information processing. The photons emitted from one interface can be fed into the another, using mirrors, reminiscent to the end-fire coupling in plasmonic systems \cite{fisher}. In this way, the nonclassical properties and entanglement of photons could be transferred to distant interfaces. This would also allow one to build a quantum network (each interface is equivalent to the node of a quantum network) at the nanoscale.}
\label {mirror}
\end{figure}

Note that, if not detected, the emitted photons are prepared in the state $|\psi\rangle_2$, which is a N00N-like state \cite{boto}. This means that the same architecture could also generate a state like $|\psi\rangle_N\equiv (|N,0\rangle + e^{\iota \phi}|0,N\rangle)/\sqrt{2}$, if one would use $N$ dots, all initially excited. In that case, the relative distance between the dots should be such that they all lie within a length scale of $\lambda_{SP}$. These photons, once radiated out, can therefore be used for metrology as can be done with N00N states \cite{dowling,taylor,zang}. These photons could also be coupled into further sets of metal-dielectric interfaces (see Fig.  \ref{mirror}). This makes our model a plausible plasmonics-based architecture for quantum information processing.

In summary, we have explored possibilities of generating nonclassical states of photons using quantum dots placed on a metal-dielectric interface. We have found that the photons in SP mode can radiate out of the interface into the radiative modes and exhibit nonclassical properties, namely, photon anticoalescence. This indicates that these photons are prepared in N00N-like path-entangled states and therefore could be used for metrology. More importantly, our proposed architecture has the potential as a building block of plasmonics-based quantum information processing.

The authors sincerely acknowledge fruitful discussion with Prof. S. Dutta Gupta, Dr. Achanta Venu Gopal and Dr. Rajesh V. Nair during the early stage of this work. The authors also thank Dr. Asoka Biswas for critical reading of the manuscript.

\bibliographystyle{elsarticl-num}

\end{document}